\documentstyle [12pt] {article}

\catcode`@=12
\begin{document}
\thispagestyle{empty}
\begin{flushright}
MAD/PH/811\\
UCRHEP-T120\\
December 1993\\
\end{flushright}
\vspace{1.0in}
\begin{center}
{\large \bf Natural Conservation of R Parity in Supersymmetry\\}
\vspace{1.0in}
V. Barger$^a$ and Ernest Ma$^{a,b}$
\vspace{0.2in}

$^a${\sl Department of Physics, University of Wisconsin, Madison, Wisconsin
53706\\}

$^b${\sl Department of Physics, University of California, Riverside,
California 92521\\}
\vspace{1.5in}
\end{center}
\begin{abstract}\
Since baryon number B and lepton number L are no longer automatically conserved
once the standard model is extended to include supersymmetry, the usually
assumed conservation of $\rm R \equiv (-1)^{2j+3B+L}$ is an imposed condition.
For a more satisfactory realization of supersymmetry, we propose
here a new model which conserves R automatically.
It is unifiable under SO(14) and has exotic fermions at the
100 GeV scale.  One important result is the enhancement of the Higgs-boson
decay rates into two gluons and into two photons by factors of 25 and 15
respectively.
\end{abstract}

\newpage
\baselineskip 24pt

In discussing supersymmetry in particle physics, it is common practice\cite{1}
to consider a quantity called R parity which is defined as
\begin{equation}
\rm R \equiv (-1)^{2j+3B+L},
\end{equation}
where j is the spin of the particle, B its baryon number and L its lepton
number.  If B and L are additively conserved, then it is obvious that
R has to be multiplicatively
conserved.  However, in the supersymmetric standard model, this is not
necessarily the case.  Consider the quark and lepton superfields.  In a
notation where only the left chiral projections are counted, they
transform under SU(3) $\times$ SU(2) $\times$ U(1) as follows:
$Q \equiv (u,d) \sim (3,2,1/6)$, $u^c \sim (\overline 3, 1, -2/3)$,
$d^c \sim (\overline 3, 1, 1/3)$, $L \equiv (\nu, e) \sim (1,2,-1/2)$,
and $e^c \sim (1,1,1)$, where the family index has been suppressed.
In addition, there must be two Higgs superfields $\Phi_{1,2}$ transforming
as $(1,2,\mp 1/2)$ respectively.  The desirable allowed terms in the
superpotential are then $\Phi_1 Q d^c$, $\Phi_2 Q u^c$, and $\Phi_1 L e^c$,
which supply the quarks and leptons with masses as the neutral scalar
components of $\Phi_{1,2}$ acquire nonzero vacuum expectation values.
However, the terms $u^c d^c d^c$, $L Q d^c$, and $L L e^c$ are also
allowed {\it a priori} and they violate the conservation of B and L.\cite{2}
Note also that $\Phi_1$ and $L$ are indistinguishable by their
transformations alone.

To obtain a realistic model, the usual procedure is to impose B and L
conservation as an extra condition.  In that case, R can be used to
distinguish particles (R = +1) from superparticles (R = $-1$).  As a
result, the lightest supersymmetric particle (LSP) is stable and
that is one of the essential features of supersymmetry upon which
experimental search strategies are based. It seems to us that it
would be far more satisfactory if B and L were automatically
conserved as in the standard model without supersymmetry.  Consider
then the conventional left-right supersymmetric extension of the
standard model.  The gauge group is now $\rm SU(3) \times SU(2)_L
\times SU(2)_R \times U(1)$.  The quarks and leptons are
\begin{eqnarray}
Q \equiv (u,d) \sim (3,2,1;1/6), &~~~& Q^c \equiv (d^c,u^c) \sim
(\overline 3, 1, 2; -1/6), \\ L \equiv (\nu, e) \sim (1,2,1;-1/2),
&~~~& L^c \equiv (e^c, \nu^c) \sim (1,1,2;1/2).
\end{eqnarray}
Hence terms involving three such superfields are not possible in
the superpotential because of gauge invariance, and the automatic
conservation of B and L appears to have been achieved.  However,
if the Higgs sector consists of only triplets, bidoublets, and
singlets, the scalar neutrinos must acquire nonzero vacuum
expectation values,
$\langle \tilde \nu^c \rangle \neq 0$, in order to
break the left-right symmetry.\cite{3}  Hence B is conserved
but not L.  [There are other
complications such as the inevitability of flavor-changing
neutral currents (FCNC) because two bidoublets are needed for
realistic quark mass matrices, as well as  the fine tuning
required to make $\langle \tilde \nu \rangle$ small and
to keep the desirable equality $\rm M_W = M_Z \cos \theta_W$
at tree level in the presence of
the $\rm SU(2)_L$ Higgs triplets.]  A recent proposal by one of
us\cite{4} also conserves B but not L, unless it is imposed.

The problem of B and L conservation in supersymmetry may be
considered also in the context of grand unification.  If SU(5)
is used, the problem persists because the $u^cd^cd^c$ term
can still come from the invariant formed with a {\bf {10}}
and two {\bf {5}$^*$} representations, and $\Phi_1$ and $L$ are
still indistinguishable as {\bf {5}$^*$}'s.  If SO(10)
is used, then the $u^cd^cd^c$ term is not allowed, and $\Phi_1$
and $\Phi_2$ belong in the {\bf {10}} whereas $L$ and $L^c$
belong in the {\bf {16}}.  This is a good solution for B
conservation as long as the exotic SU(3) triplets in
the {\bf {10}} are made very heavy.\cite{5}  However, as
SO(10) contains $\rm SU(3) \times SU(2)_L \times SU(2)_R \times
U(1)$, the details of the symmetry breaking still require
nonzero vacuum expectation values for $\rm SU(2)_R$ and
$\rm SU(2)_L$ doublets.  Hence L conservation is broken
either spontaneously\cite{3} if the only such doublets are
leptons, or explicitly\cite{4} as well if there are additional
Higgs doublet superfields and no extra discrete symmetry
is assumed to distinguish them from the leptons.

Consider now the addition of another U(1) factor which contains
electric charge but under which $Q$, $Q^c$, $L$, and $L^c$
transform trivially.  We also add the following superfields:
\begin{eqnarray}
x \sim (3,1,1;-1/3,1), &~~~& x^c \sim (\overline 3,1,1;1/3,-1), \\
N \sim (1,1,1;-1,1), &~~~& N^c \sim (1,1,1;1,-1),
\end{eqnarray}
\begin{equation}
\Phi_{12} \equiv \left( \begin{array} {c@{\quad}c} \overline
{\phi_1^0} & \phi_2^+ \\ -\phi_1^- & \phi_2^0 \end{array}
\right) \sim (1,2,2;0,0),
\end{equation}
and
\begin{equation}
\Phi_3 \equiv (\phi_3^+,\phi_3^0) \sim (1,2,1;-1/2,1), ~~~~ \Phi_4
\equiv (\overline {\phi_4^0}, -\phi_4^-) \sim (1,1,2;1/2,-1).
\end{equation}
Note that $\Phi_3$ and $\Phi_4$ transform differently from $L$
and $L^c$ under the extra U(1).  Note also that the above particle
content is not anomaly-free.  In Ref. [4], two
$\rm SU(2)_R$ doublets are used so that the structure
is anomaly-free, but then since one of the new superfields transforms
identically as $L^c$, the conservation of lepton number cannot be
maintained without being imposed.  Here the allowed Yukawa
terms are
\begin{eqnarray}
\Phi_{12} Q Q^c &=& \overline {\phi_1^0} d d^c + \phi_1^- u d^c
+ \phi_2^0 u u^c - \phi_2^+ d u^c, \\ \Phi_{12} L L^c &=&
\overline {\phi_1^0} e e^c + \phi_1^- \nu e^c
+ \phi_2^0 \nu \nu^c - \phi_2^+ e \nu^c,
\end{eqnarray}
and
\begin{eqnarray}
\Phi_3 Q x^c = \phi_3^0 u x^c - \phi_3^+ d x^c, &~~~&
\Phi_4 x Q^c = \overline {\phi_4^0} x u^c + \phi_4^- x d^c, \\
\Phi_3 L N^c = \phi_3^0 \nu N^c - \phi_3^+ e N^c, &~~~&
\Phi_4 N L^c = \overline {\phi_4^0} N \nu^c + \phi_4^- N e^c.
\end{eqnarray}
Hence B and L are automatically conserved with B = 1/3 for $Q$ and
$x$, B = $-1/3$ for $Q^c$ and $x^c$, L = 1 for $L$ and $N$, and
L = $-1$ for $L^c$ and $N^c$, where the singlet quark $x$ has
charge 2/3 and the singlet lepton $N$ is neutral.  The Higgs
superfields $\Phi_{12}$, $\Phi_3$, and $\Phi_4$ have B = L = 0.
This assignment is automatic without the need of any extra
imposed condition because
they are in representations different from $L$, $L^c$, $N$, and
$N^c$.  As a result, the spontaneous breaking of the gauge
symmetry through the nonzero vacuum expectation values of
$\Phi_{12}$, $\Phi_3$, and $\Phi_4$ will not violate the
conservation of B and L.  However, $\Phi_3$ and $\Phi_4$
generate nonvanishing axial-vector anomalies as already
mentioned and we should think about how they are to be
canceled.

Consider the gauge group SO(10) $\times$ U(1).  It is obvious
that $Q$, $Q^c$, $L$, and $L^c$ are in the ({\bf 16},0)
representation whereas $\Phi_{12}$ is in the ({\bf 10},0).
It is thus natural to assume that $\Phi_3$ is in the ({\bf 16},1)
and $\Phi_4$ is in the ({\bf 16},$-1$).  The extra color triplets
and singlets in the ({\bf 16},$\pm 1$) will then render the theory
anomaly-free.  The fermions in these superfields will also have
masses at the electroweak energy scale because they couple to
$\Phi_{12}$ in analogy to the usual quarks and leptons. As for the
singlet quarks $x$ and $x^c$ and the singlet leptons $N$ and
$N^c$, although they do not generate anomalies at the $\rm SU(3)
\times SU(2) \times SU(2) \times U(1) \times U(1)$ level, they
can also be considered as belonging to the ({\bf 10},$\pm 1$)
and ({\bf 120},$\pm 1$) representations of SO(10) $\times$ U(1)
respectively.  Later on, we will show that it is natural to
extend the gauge group further to SO(14), but now let us return
to the interaction structure of our model at low energies.

Since $x$ and $x^c$ are singlets, there is an allowed
gauge-invariant mass term $xx^c$.  Let $\langle \phi_{1,2,3,4}^0
\rangle = v_{1,2,3,4}$, then the 6 $\times$ 6 mass matrix
linking $(u,x)$ with $(u^c,x^c)$ is given by
\begin{equation}
{\cal M}_{ux} = \left[ \begin{array} {c@{\quad}c} v_2 v_1^{-1}
{\cal M}_d & {\cal M}_3 \\ {\cal M}_4 & {\cal M}_x \end{array}
\right],
\end{equation}
where the 3 $\times$ 3 mass matrices ${\cal M}_d$ and
${\cal M}_x$ can be chosen to be diagonal, with  ${\cal M}_3$
and ${\cal M}_4$ proportional to $v_3$ and $v_4$
respectively.  The mixing of $u$ and $x$ is determined by the
matrix
\begin{equation}
{\cal M}_{ux} {\cal M}_{ux}^\dagger = \left[ \begin{array}
{c@{\quad}c} v_2^2 v_1^{-2} {\cal M}_d {\cal M}_d^\dagger +
{\cal M}_3 {\cal M}_3^\dagger &
v_2 v_1^{-1} {\cal M}_d {\cal M}_4^\dagger + {\cal M}_3
{\cal M}_x^\dagger \\ v_2 v_1^{-1} {\cal M}_4 {\cal M}_d^\dagger
+ {\cal M}_x {\cal M}_3^\dagger & {\cal M}_4 {\cal M}_4^\dagger
+ {\cal M}_x {\cal M}_x^\dagger \end{array} \right],
\end{equation}
and since $v_3$ breaks $\rm SU(2)_L$ but $v_4$ breaks $\rm SU(2)_R$,
it is clear that $u-x$ mixing is very small and can be safely
neglected.  On the other hand, the mass matrix for the $u$
quarks is given by
\begin{eqnarray}
{\cal M}_u {\cal M}_u^\dagger &=& v_2^2 v_1^{-2} {\cal M}_d
{\cal M}_d^\dagger + {\cal M}_3 {\cal M}_3^\dagger
- (v_2 v_1^{-1} {\cal M}_d {\cal M}_4^\dagger
+ {\cal M}_3 {\cal M}_x^\dagger) \nonumber \\ &\times&
({\cal M}_4 {\cal M}_4^\dagger + {\cal M}_x {\cal M}_x^\dagger)^{-1}
(v_2 v_1^{-1}{\cal M}_4 {\cal M}_d^\dagger +
{\cal M}_x {\cal M}_3^\dagger),
\end{eqnarray}
which is in general nondiagonal and can easily be
phenomenologically correct even if $u-x$ mixing is very small.
The mixing of $u^c$ and $x^c$ is determined by the matrix
${\cal M}_{ux}^\dagger {\cal M}_{ux}$ and can be quite large
because ${\cal M}_4$ and ${\cal M}_x$ may be comparable in
magnitude.
However, because $u^c$ and $x^c$ transform identically
under the standard SU(2) $\times$ U(1), their coupling
to the Z boson remains diagonal.  Nevertheless, there
are flavor-changing neutral currents in the $u$ sector
due to the exchange of the Z$'$ boson from $\rm SU(2)_R
\times U(1)$ breaking and that of neutral Higgs bosons.
One interesting implication\cite{4} is the
possibility of the decay $t \rightarrow c$ + Higgs boson,
which may be observable once a large enough sample of
$t$'s are available experimentally.  We will come back
to this point later.

In the leptonic sector, the analogous mass
matrix linking $(\nu,N)$ with $(\nu^c,N^c)$ is given by
\begin{equation}
{\cal M}_{\nu N} = \left[ \begin{array} {c@{\quad}c}
v_2 v_1^{-1} {\cal M}_\ell & {\cal M}_3' \\ {\cal M}_4'
& {\cal M}_N \end{array} \right],
\end{equation}
where ${\cal M}_\ell$ is the 3 $\times$ 3 charged-lepton
mass matrix.  However, very delicate fine tuning would then
be required to obtain the necessary small neutrino masses.
A more natural solution is to allow Majorana masses for
$N$ and $N^c$ which can come from the vacuum expectation
values of superfields transforming as (1,1,1;2,$-2$) and
(1,1,1;$-2$,2) respectively.  [Actually their presence
serves a dual purpose.  Without them and with only
$\Phi_{12}$, $\Phi_3$, and $\Phi_4$, a linear combination
of the two U(1) factors would stay unbroken in addition
to the electromagnetic U(1).]  The induced Majorana mass
matrix for $\nu^c$ is then given by ${{\cal M}_4'}^\dagger
{\cal M}_{\rm eff}^{-1} {\cal M}_4'$, where
${\cal M}_{\rm eff}$ is determined by the heavy $(N,N^c)$
mass submatrix, and can have large mass eigenvalues,
say of order $10^2$ GeV.  The Majorana neutrino mass
matrix ${\cal M}_\nu$ gets two see-saw contributions:
one from the above-mentioned $\nu^c$ masses and is
given by $v_2^2 v_1^{-2} {\cal M}_\ell {\cal M}_{\nu^c}^{-1}
{\cal M}_\ell^\dagger$, the other coming from ${\cal M}_3'$
and a different ${\cal M}_{\rm eff}$.  Both effects are highly
suppressed, hence the observed neutrinos have naturally
small Majorana masses in this model.  Additive lepton number
is now broken, but a multiplicative lepton number is still
conserved: $L$, $L^c$, $N$, and $N^c$ are odd and all
other superfields are even.  Hence the automatic
conservation of R parity remains valid.

The spontaneous breaking of the $\rm SU(3) \times SU(2)_L
\times SU(2)_R \times U(1) \times U(1)$ gauge symmetry is
accomplished in this model first by the (1,1,1;2,$-2$) and
(1,1,1;$-2$,2) singlets which reduce the two U(1) factors
into one.  Then $\Phi_4$ breaks $\rm SU(2)_L \times
SU(2)_R \times U(1)$ down to the standard $\rm SU(2)_L \times
U(1)$ which is in turn broken down to the electromagnetic
U(1) by $\Phi_{12}$ and $\Phi_3$.  The superpotential
consisting of $\Phi_{12}$, $\Phi_3$, and $\Phi_4$ is given
by
\begin{equation}
W = m {\rm det} \Phi_{12} + f \tilde \Phi_3^\dagger \Phi_{12}
\tilde \Phi_4,
\end{equation}
where
\begin{equation}
\tilde \Phi_3 \equiv i \sigma_2 \Phi_3^* = \left( \begin{array}
{c} \overline {\phi_3^0} \\ - \phi_3^- \end{array} \right),
\end{equation}
and we have redefined $\Phi_4$ as $\tilde \Phi_4$ so that both
$\Phi_3$ and $\Phi_4$ now have scalar components denoted by
$(\phi_3^+,\phi_3^0)$ and $(\phi_4^+,\phi_4^0)$ respectively
in accordance with the notation $\Phi_1 = (\phi_1^+,\phi_1^0)$
and $\Phi_2 = (\phi_2^+,\phi_2^0)$ for the scalar components of
$\Phi_{12}$.  The soft terms of the Higgs potential, including
those which break the supersymmetry, are given by
\begin{eqnarray}
V_{\rm soft} &=& m^2 {\rm Tr} (\Phi_{12}^\dagger \Phi_{12}) +
mB ({\rm det} \Phi_{12} + {\rm det} \Phi_{12}^\dagger) \nonumber \\
&+& m_3^2 \Phi_3^\dagger
\Phi_3 + m_4^2 \Phi_4^\dagger \Phi_4 + fA (\tilde \Phi_3^\dagger
\Phi_{12} \tilde \Phi_4 + \tilde \Phi_4^\dagger \Phi_{12}^\dagger
\tilde \Phi_3).
\end{eqnarray}
We look for a solution in which $v_{1,2,3}$ are small compared
to $v_4$.  At the electroweak energy scale, the Higgs sector
may reduce to three doublets, two doublets, or one doublet.  The
three-doublet case occurs only if $f^2 = g^2/2$ and both $mB$
and $fAv_4$ are of order $(100~GeV)^2$.  In the
two-doublet case, consisting of $\Phi_1$ and a linear
combination of $\Phi_2$ and $\Phi_3$, the minimal supersymmetric
standard model (MSSM) is obtained in the limit
$f=0$.  If $f \neq 0$, the two doublets will not have the
couplings of the MSSM and a different mass spectrum will be
found.\cite{6}  In the one-doublet case, we have of course
only one physical Higgs boson as in the standard model.
Specifically, it is given here by
\begin{equation}
h = {{\sqrt 2 (v_1 {\rm Re} \phi_1^0 + v_2 {\rm Re} \phi_2^0 +
v_3 {\rm Re} \phi_3^0)} \over {\sqrt {v_1^2 + v_2^2 + v_3^2}}}.
\end{equation}
Whereas $h$ couples to $dd^c$ according to ${\cal M}_d$ as in
the standard model, it couples to $uu^c$ according to $v_2v_1^{-1}
{\cal M}_d$ and $ux^c$ according to ${\cal M}_3$, hence there
are FCNC effects in the $u$ sector due to the exchange of $h$.
[Recall there can be large mixing between $u^c$ and $x^c$ in this
model.]

Consider now the decay $t \rightarrow c + h$.  The coupling is
equal to $(\xi g/2)(m_t/{\rm M_W})$, where $\xi$ is a suppression
factor due to mixing.  Hence
\begin{equation}
{{\Gamma (t \rightarrow c + h)} \over {\Gamma (t \rightarrow b +
{\rm W})}} = \xi^2 \left( 1 - {m_h^2 \over m_t^2} \right)^2 \left(
1 - {{\rm M_W}^2 \over m_t^2} \right)^{-2} \left( 1 + {{2 {\rm M_W}^2}
\over m_t^2} \right)^{-1}.
\end{equation}
Since the value of $\xi$ is unconstrained by present experimental
data, the above ratio may be substantial.  Once produced, the
Higgs boson $h$ will decay into $b \overline b$.
The background to $t \rightarrow c + h$ is thus mainly $t \rightarrow
b + {\rm W}$, where the W decays into $c \overline b$.  However,
the latter is suppressed by $|V_{cb}|^2 \sim 2 \times 10^{-3}$
and the $b \overline b$ invariant mass will not peak at $m_h$.
In a hadron collider such as the Tevatron at Fermilab, a
$t \overline t$ pair can be produced if kinematically allowed,
then if $\overline t \rightarrow \overline b + {\rm W}$,
where the W decays into
an electron (or a muon) and its antineutrino, the decay $t
\rightarrow c + h$, where $h \rightarrow b \overline b$,
may have a chance of being observed through the use of vertex
detectors.

Let us return now to the exotic fermions contained in the
({\bf 16},$\pm 1$) of SO(10) $\times$ U(1).  They acquire masses
through $\Phi_{12}$ and must therefore not be very heavy.  They
also interact with the singlet quarks $x$ and $x^c$ contained in
the ({\bf 10},$\pm 1$) representations.  Under SO(10) $\times$
U(1), there are in fact only four types of Yukawa terms:
({\bf 16},0)({\bf 16},0)({\bf 10},0),
({\bf 16},1)({\bf 16},$-1$)({\bf 10},0), and
({\bf 16},0)({\bf 16},$\pm 1$)({\bf 10},$\mp 1$).  Assuming
that at the $\rm SU(3) \times SU(2)_L \times SU(2)_R \times U(1)
\times U(1)$ level, the ({\bf 10},0) contains only $\Phi_{12}$
and the ({\bf 10},$\pm 1$) contain only $x$, $x^c$, and
\begin{equation}
y \sim (3,1,1;-1/3,-1),~~~~~~y^c \sim (\overline 3,1,1;1/3,1),
\end{equation}
then every B and L assigment is uniquely determined.  The complete
list is given in Table 1.  Note that because of the structure of
this model, the new particles have unusual baryon and lepton numbers.
Note also that the ({\bf 10},$\pm 1$) fermions have R = +1, whereas
the ({\bf 16}, $\pm 1$) fermions have R = $-1$.  The decay products
of the latter must then always include the LSP, which we will
choose for convenience to be the photino $\tilde \gamma$.

Consider now the SU(3)-triplet fermion which has electric
charge 5/3.  It has a Yukawa coupling to $Lx^c$.  Hence it will
decay into $e^+ u \tilde \gamma$ through $u^c-x^c$ mixing and
the exchange of the heavy squark $\tilde u^c$.  The absence of
such a signal above background at the Tevatron so far suggests
that it has a mass greater than 100 GeV.  On the other
hand, the magnitude of
its Yukawa coupling is related to ${\cal M}_3$ of Eq. (12)
in the SO(10) $\times$ U(1) limit and is likely to be rather small
for the physical $u$ and $c$ quarks.  Therefore, we expect a
decay rate orders of magnitude smaller than that of
ordinary heavy quarks such as the $t$ which can decay into a
physical W boson + another quark.  Consequently, a bound state
of these exotic SU(3)-triplet fermions may exist up to a much
higher mass than the usual quarkonia.\cite{7}  The scalar ground
state will decay dominantly into two gluons, whereas the branching
fraction into two photons is given by
$(5/3)^4(3/8)(\alpha/\alpha_s)^2$ which is about 2\%.

Another important consequence of the ({\bf 16},$\pm 1$) fermions
is their contribution to the effective two-gluon and two-photon
couplings of the Higgs boson $h$.  These couplings are absent at
tree level but are nonzero to one loop where all particles which
couple to $h$ will contribute, depending of course on whether they
also couple to gluons or to photons.  In this model, $\Gamma (h
\rightarrow gg)$ is enhanced over that of the standard model by
a factor of roughly $(1+4)^2 = 25$ because there are now four more
heavy quarks.  Assume for illustration that $m_h = 90$ GeV, then the
cross section for $p \overline p \rightarrow h$ + anything at a
center-of-mass energy of 2 TeV is about 35 pb.\cite{8}  Similarly,
$\Gamma (h \rightarrow \gamma \gamma)$ is enhanced by a factor of
roughly 15, and $B(h \rightarrow \gamma \gamma)$ is about
$9 \times 10^{-3}$.\cite{9}  These two large enhancements
make it much easier for $h$ to be discovered at the Tevatron
with a signal of about 0.3 pb above a background\cite{8} of
about 0.1 pb for this value of $m_h$.  Note that the heavy-quark
contribution to the background is negligible when the invariant
mass of the photon pair is much less than that of the quark
pair.\cite {10}  Details of this and other
phenomenological implications will be given elsewhere.

If further unification is desired beyond SO(10) $\times$ U(1), the
natural choice is SO(14).  Consider the latter's SO(10) $\times$
SO(4) decomposition.  It is clear that the spinorial {\bf 64}
representation of SO(14) splits up into four {\bf 16} representations
of SO(10), and each is a component of the {\bf 4} representation of
SO(4).  Since the U(1) decomposition of the latter has charges 1, 0, 0,
$-1$, it is also clear that our ({\bf 16},0) and ({\bf 16},$\pm 1$)
representations can be accommodated.  Similarly, the ({\bf 10},$\pm 1$)
and ({\bf 120},$\pm 1$) representations are accommodated in the
product {\bf 64} $\times$ {\bf 64} of SO(14).

In conclusion, we have shown how R parity can be automatically
conserved in a realistic model of supersymmetry.  It is based on the
gauge group $\rm SU(3) \times SU(2)_L \times SU(2)_R \times U(1)
\times U(1)$ and is unifiable under SO(14).  A necessary condition
is to make sure that the Higgs superfields needed for the
spontaneous breaking of the gauge symmetry are in representations
different from those of the leptons.  We want to use only Higgs
superfields which are doublets or singlets under the standard
$\rm SU(2)_L \times U(1)$ so that the tree-level
equality $\rm M_W = M_Z cos \theta_W$ can be maintained, hence
the choice of $\Phi_{12}$, $\Phi_3$, and $\Phi_4$.  Motivated
by the necessity of anomaly cancellation and the possibility of
grand unification, we put $\Phi_3$ and $\Phi_4$
in the ({\bf 16},$\pm 1$)
representations of SO(10) $\times$ U(1).  The fermions
contained therein must not be very heavy because they get their
masses through $\Phi_{12}$.  They also interact with the singlet
superfields $x$ and $x^c$ which are introduced to mix with
$u$ and $u^c$ to obtain a realistic ${\cal M}_u$ despite having
only one $\Phi_{12}$.  Consequently, every B and L assignment is
uniquely determined in this model, as shown in Table 1.
Two particularly interesting
phenomenological implications are the possibility
of heavy bound states
of exotic color-triplet fermions with significant branching
fractions into two photons and that of greatly enhanced
two-gluon and two-photon couplings of the Higgs boson $h$.
\newpage
\begin{center} {ACKNOWLEDGEMENT}
\end{center}

We thank A. Stange and S. Willenbrock for discussions of the
results of Ref. [8].
This work was supported in part by the U. S. Department of Energy
under contracts No. DE-AC02-76ER00881 and No. DE-AT03-87ER40327,
by the Texas National Laboratory Research Commission, and by the
University of Wisconsin Research Committee with funds granted by
the Wisconsin Alumni Research Foundation.
\vspace{1.0in}

\bibliographystyle{unsrt}

\begin{thebibliography}{99}
\bibitem{1} For a review, see for example R. Barbieri, {\em La
Rivista del Nuovo Cimento}, Vol. 11, N. 4 (1988).
\bibitem{2} There are numerous papers on this subject.  A partial
list includes L.J. Hall and M. Suzuki, {\em Nucl. Phys.} {\bf B231},
419 (1984); I.H. Lee, {\it ibid.} {\bf B246}, 120 (1984); J. Ellis
{\it et al.}, {\em Phys. Lett.} {\bf 150B}, 142 (1985); G.G. Ross
and J.W.F. Valle, {\it ibid.} {\bf 151B}, 375 (1985); S. Dawson,
{\em Nucl. Phys.} {\bf B261}, 297 (1985); R. Barbieri and A. Masiero,
{\it ibid.} {\bf B267}, 679 (1986); R. N. Mohapatra, {\em Phys. Rev.}
{\bf D34}, 3457 (1986);
S. Dimopoulos and L.J. Hall, {\em Phys. Lett} {\bf B207}, 210 (1987);
V. Barger, G.F. Giudice, and T. Han, {\em Phys. Rev.} {\bf D40},
2987 (1989); E. Ma and P. Roy, {\it ibid.} {\bf D41}, 988 (1990);
E. Ma and D. Ng, {\it ibid.} {\bf D41}, 1005 (1990);
S. Dimopoulos {\it et al.}, {\it ibid.} {\bf D41}, 2099 (1990);
\bibitem{3} R. Kuchimanchi and R. N. Mohapatra, {\em Phys. Rev.}
{\bf D48}, 4352 (1993).
\bibitem{4} E. Ma, Univ. of Wisconsin, Madison Report No. MAD/PH/792 (Sep 93).
\bibitem{5} K.S. Babu and S.M. Barr, Bartol Res. Inst. Report No.
BA-93-26 (May 93).
\bibitem{6} E. Ma and D. Ng, Univ. of Calif., Riverside Report No.
UCRHEP-T107 (1993); T. V. Duong and E. Ma, {\em Phys. Lett.} {\bf B316},
307 (1993).
\bibitem{7} J.H. K$\ddot{\rm u}$hn and P.M. Zerwas, {\em Phys. Rep.} {\bf 167},
321 (1988); V. Barger {\it et al.}, {\em Phys. Rev.} {\bf D35}, 3366 (1987).
\bibitem{8} A. Stange, W. Marciano, and S. Willenbrock, Fermilab Report No.
FERMILAB-PUB-93/142-T (1993).
\bibitem{9} V. Barger {\it et al.}, Univ. of Wisconsin, Madison Report No.
MAD/PH/749, {\em Phys. Rev.} {\bf D} (in press).
\bibitem{10} D.A. Dicus and S. Willenbrock, {\em Phys. Rev.} {\bf D37}, 1801
(1988).
\end{thebibliography}

\newpage
\begin{table}
\begin{center}
\begin{math}
\begin{array} {|c|c|c|c|c|c|} \hline
{}~ & \rm Representation & \rm Charge & \rm B & \rm L & \rm R_f \\
\hline
Q & (3,2,1;1/6,0) & (2/3,-1/3) & 1/3 & 0 & + \\
Q^c & (\overline 3,1,2;-1/6,0) & (1/3,-2/3) & -1/3 & 0 & + \\
L & (1,2,1;-1/2,0) & (0,-1) & 0 & 1 & + \\
L^c & (1,1,2;1/2,0) & (1,0) & 0 & -1 & + \\
\hline
N & (1,1,1;-1,1) & 0 & 0 & 1 & + \\
N^c & (1,1,1;1,-1) & 0 & 0 & -1 & + \\
x & (3,1,1;-1/3,1) & 2/3 & 1/3 & 0 & + \\
x^c & (\overline 3,1,1;1/3,-1) & -2/3 & -1/3 & 0 & + \\
y & (3,1,1;-1/3,-1) & -4/3 & -2/3 & 1 & + \\
y^c & (\overline 3,1,1;1/3,1) & 4/3 & 2/3 & -1 & + \\
\hline
\Phi_{12} & (1,2,2;0,0) & (1,0,0,-1) & 0 & 0 & - \\
\Phi_3 & (1,2,1;-1/2,1) & (1,0) & 0 & 0 & - \\
\Phi_4 & (1,1,2;1/2,-1) & (0,-1) & 0 & 0 & - \\
\hline
L_- & (1,2,1;-1/2,-1) & (-1,-2) & -1 & 1 & - \\
L_-^c & (1,1,2;1/2,1) & (2,1) & 1 & -1 & - \\
Q_+ & (3,2,1;1/6,1) & (5/3,2/3) & 1/3 & -1 & - \\
Q_+^c & (\overline 3,1,2;-1/6,-1) & (-2/3,-5/3) & -1/3 & 1 & - \\
Q_- & (3,2,1;1/6,-1) & (-1/3,-4/3) & -2/3 & 0 & - \\
Q_-^c & (\overline 3,1,2;-1/6,1) & (4/3,1/3) & 2/3 & 0 & - \\
\hline
\end{array}
\end{math}
\end{center}
\caption {Particle content of this model under $\rm SU(3) \times
SU(2)_L \times SU(2)_R \times U(1) \times U(1)$ and the associated
electric charge, baryon number, lepton number, and R parity of the
fermions.}
\end{table}

\end{document}